\documentclass[12pt]{article}
\usepackage{graphicx}
\usepackage{scicite}
\usepackage{amsmath}
\usepackage{amssymb}
\usepackage{float}
\usepackage{bm,times}
\usepackage{graphicx}
\newcommand{\be}{\begin{equation}}
\newcommand{\ee}{\end{equation}}
\newcommand{\bea}{\begin{eqnarray}}
\newcommand{\eea}{\end{eqnarray}}
\newcommand{\ek}{\epsilon_{\mathbf{k}}}

\newcommand{\Ek}{E_{\mathbf{k}}}

\newcommand{\sumk}{\sum_{\mathbf{k}}}

\newcommand{\Omegaq}{\Omega_{\mathbf{q}}}

\newenvironment{sciabstract}{%
\begin{quote} \bf}
{\end{quote}}

\newcounter{lastnote}
\newenvironment{scilastnote}{%
\setcounter{lastnote}{\value{enumiv}}%
\addtocounter{lastnote}{+1}%
\begin{list}%
{\arabic{lastnote}.}
{\setlength{\leftmargin}{.22in}}
{\setlength{\labelsep}{.5em}}}
{\end{list}}

\title{Heat Capacity of a Strongly-Interacting Fermi Gas}

\author{J. Kinast,$^1$ A. Turlapov,$^1$  J. E. Thomas,$^1$$^{\ast}$
\\
Qijin Chen,$^2$ Jelena Stajic,$^2$ and K. Levin$^2$\\
\\
\normalsize{$^1$Physics Department, Duke University, Durham, North
Carolina 27708-0305, USA}
\\
\normalsize{$^2$James Franck Institute and Department of Physics,
University of Chicago,}\\
\normalsize{5640 South Ellis Avenue, Chicago, Illinois 60637, USA}\\
\\
\normalsize{$^\ast$To whom correspondence should be addressed;
E-mail:  jet@phy.duke.edu.}
\\
\normalsize{  }\\
\normalsize{(Submitted 09 November, 2004)}\\
\normalsize{}}
\date{}

\begin{document}

\baselineskip24pt

\maketitle

\begin{sciabstract}

  We have measured the heat capacity of an
  optically-trapped, strongly-interacting Fermi gas of atoms.
  A precise input of energy to the gas is followed by
  single-parameter thermometry, which  determines the
empirical temperature parameter $\tilde{T}$ of the gas cloud. Our
measurements reveal a clear transition in the heat capacity.
 The energy and the spatial profile of the gas
are computed using a theory of the crossover from Fermi to Bose
superfluids at finite temperature. The theory calibrates
$\tilde{T}$, yields excellent agreement with the data, and
predicts the onset of superfluidity at the observed transition
point.

\end{sciabstract}

Strongly-interacting, degenerate atomic Fermi
gases~\cite{OHaraScience}  provide a paradigm for strong
interactions in nature~\cite{AmScientist}. In all strongly
interacting Fermi systems, the zero-energy scattering length is
large compared to the interparticle spacing, producing universal
behavior~\cite{Heiselberg,HoUniversalThermo}. Predictions of
universal interactions and effective field theories in nuclear
matter~\cite{Heiselberg,Baker,Carlson,Strinati} are tested by
measurements of the interaction
energy~\cite{OHaraScience,MechStab,SalomonBEC,Grimmbeta}.
Anisotropic expansion of strongly-interacting Fermi
gases~\cite{OHaraScience} is analogous to the ``elliptic flow" of
a quark-gluon plasma~\cite{AmScientist}. High temperature
superfluidity has been predicted
~\cite{Houbiers,CombescotHighTc,Holland,Timmermans,Griffin,Stajic}
in strongly-interacting Fermi gases, which can be used to test
theories of high temperature superconductivity~\cite{Levin}.
Microscopic evidence for superfluidity has been obtained by
observing the pairing of fermionic
atoms~\cite{Jincondpairs,Ketterlecondpairs,GrimmGap}. Macroscopic
evidence arises in anisotropic expansion~\cite{OHaraScience} and
in collective excitations~\cite{Kinast,Bartenstein,KinastMagDep}.

In superconductivity and superfluidity, measurements of the heat
capacity have played an exceptionally important role in
determining phase transitions~\cite{LondonBEC} and in
characterizing the nature of bosonic and fermionic excitations. We
report on the measurement of the heat capacity for a
strongly-interacting Fermi gas of $^6$Li atoms, confined in an
optical trap. Our experiments~\cite{KinastScience} examine the
fundamental thermodynamics of the gas.

Thermodynamical properties of the BCS-BEC crossover system are
computed~\cite{SupportOnline} using a consistent many-body
theory~\cite{ChenScience,Chen4} based on the conventional mean
field state~\cite{Leggett}. BCS-BEC crossover refers to the smooth
change from the Bardeen-Cooper-Schrieffer superfluidity of
fermions to the Bose-Einstein condensation of dimers, by varying
the strength of the pairing interaction (for example, by tuning a
magnetic field). The formalism of Ref.~\cite{Stajic,Levin,Chen4}
was applied recently~\cite{Torma} to explain radio frequency
measurements of the gap~\cite{GrimmGap}. The theory contains two
contributions to the entropy and energy arising from fermionic and
bosonic excitations. The latter are associated principally with
excited pairs of fermions (Cooper pairs at finite momentum). In
this model, there is no direct boson-boson coupling, and
fermion-boson interactions are responsible for the vanishing of
the pair chemical potential $\mu_{pair}$ in the superfluid
regions.  The vanishing of $\mu_{pair}$ implies that, within a
trap, the associated low temperature power laws in the entropy and
energy are the same as those of the homogeneous system
\cite{Carr}.  This is to be contrasted with models which involve
noninteracting bosons and fermions \cite{Williams}. Clearly, our
BCS-like ground state ansatz will be inapplicable at some point
when the fermionic degrees of freedom have completely disappeared,
and the gas is deep in the BEC regime, where the power laws
associated with true, interacting bosons are expected \cite{Carr}.
In that case, direct inter-boson interactions must be accounted
for and they will alter the collective mode
behavior~\cite{Stringariosc}. However, on the basis of collective
mode experiments \cite{Kinast,Bartenstein,KinastMagDep} and their
theoretical interpretation \cite{Tosi,Heiselbergosc}, one can
argue that the BCS-like ground state appears appropriate in the
near resonance, unitary regime. The thermodynamic quantities
within the trap are computed using previously calculated
profiles~\cite{LevinDensity} of the various energy gaps and the
particle density as a function of the radius.

Unlike the weak coupling BCS limit, the pairing gap in the unitary
regime is very large. Well below the superfluid transition
temperature $T_c$, fermions are paired over much of the trap, and
unpaired fermions are present only at the edges of the trap. These
unpaired fermions tend to dominate the thermodynamics associated
with the fermionic degrees of freedom, and lead to a higher (than
linear) power law in the temperature ($T$) dependence of entropy.
The contribution from finite momentum Cooper pairs leads to a
$T^{3/2}$ dependence of the entropy on temperature. Both bosonic
and fermionic contributions are important at low $T$.

An important feature of these fermionic superfluids is that pair
formation occurs at a higher temperature $T^*$ than the
temperature $T_c$ where pairs condense. At temperatures $T>T^*$,
the entropy approaches that of the noninteracting gas. For
$T_c<T<T^*$, the attraction is strong enough to form quasi-bound
(or preformed) pairs which are reflected in the thermodynamics. At
these temperatures, a finite energy, i.e., the pseudogap, is
needed to create single fermion excitations
~\cite{Chen4,Levin,Stajic}. Interestingly, in the unitary regime,
both $T^*$ and $T_c$ are large fractions of the Fermi temperature
$T_F$, signifying high temperature pair formation and very high
 temperature superfluidity.

We prepare a degenerate, unitary Fermi gas comprising a 50-50
mixture of the two lowest spin states of $^6$Li atoms near a
Feshbach resonance. To cool the gas, we use forced evaporation at
a bias magnetic field of 840 G in an ultrastable CO$_2$ laser
trap~\cite{OHaraScience,AmScientist,SupportOnline}.  After cooling
well into the degenerate regime, energy is precisely added to the
trapped gas at fixed atom number, as described below.  The gas is
then allowed to thermalize for 0.1 s before being released from
the trap and imaged at 840 G after 1 ms of expansion to determine
the number of atoms and the temperature parameter $\tilde{T}$. For
our trap the total number of atoms is $N=2.2(0.3)\times 10^5$. The
corresponding noninteracting gas Fermi temperature is
$T_F=(3N)^{1/3}\hbar\bar{\omega}/k_B\simeq 2.5\,\mu$K, small
compared to the final trap depth of $U_0/k_B=35\,\mu$K.

Energy is precisely added to the trapped gas at fixed atom number
by releasing the cloud from the trap and permitting it to expand
for a short time $0\leq t_{heat}\leq 460\,\mu$s after which the
gas is recaptured. Even for the strongly-interacting gas, the
energy input is well-defined for very low initial temperatures,
where both the equation of state and the expansion dynamics are
known. During the times $t_{heat}$ used in the experiments, the
axial size of the gas changes negligibly, while transverse
dimensions expand by a factor $b_\perp(t_{heat})$. Hence, the mean
harmonic trapping potential energy $\langle U_{HO}\rangle$ in each
of the two transverse directions increases by a factor
$b_\perp^2(t_{heat})$.

The initial potential energy is readily determined at zero
temperature from the equation of state of the gas,
$(1+\beta)\epsilon_F(\mathbf{x})+U_{HO}(\mathbf{x})=\mu_0$
~\cite{OHaraScience,MechStab}, where $\epsilon_F(\mathbf{x})$ is
the local Fermi energy, $\beta$ is the unitary gas parameter
~\cite{OHaraScience,Heiselberg,MechStab,Carlson,Strinati},  and
$\mu_0$ is the global chemical potential. This equation of state
is supported by low temperature studies of the breathing
mode~\cite{Kinast,KinastMagDep,Stringariosc,Heiselbergosc} and the
spatial profiles~\cite{OHaraScience,Carlson,LevinDensity}. It is
equivalent to that of a harmonically trapped noninteracting gas of
particles with an effective mass ~\cite{Baker}, which in our
notation is $m^*=m/(1+\beta)$, where $m$ is the bare fermion mass.
The mean potential energy is half of the total energy, because the
gas behaves as a harmonic oscillator. As $\beta
<0$~\cite{Carlson,Strinati}, $m^*>m$, so that the effective
oscillation frequencies and the chemical potential are simply
scaled down, i.e., $\mu_0=k_BT_F\sqrt{1+\beta}$
~\cite{OHaraScience,MechStab}. The total energy at zero
temperature, which determines the energy scale, is therefore
\begin{equation}
E_0=\frac{3}{4}N\mu_0=\frac{3}{4}Nk_BT_F\sqrt{1+\beta}.
\label{eq:E0}
 \end{equation}
 For each direction, the initial potential energy at zero temperature is
 $E_0/6$. Then, the total energy of the gas after heating is given
 by,
\begin{equation}
E(t_{heat})=\eta\,E_0\left[\,\frac{2}{3}+\frac{1}{3}\,
  b_\perp^2(t_{heat})\right],
\label{eq:energy}
\end{equation}
neglecting trap anharmonicity~\cite{SupportOnline}. Here, $\eta$
is a correction factor arising from the finite temperature of the
gas prior to the energy input. For the strongly-interacting gas,
the initial reduced temperature is very low. We assume that it is
 $\simeq \tilde{T}=0.04$, where $\tilde{T}$ is measured and
calibrated as described below. Assuming a Sommerfeld correction
then yields $\eta_{int} \simeq 1+ 2\pi^2 \tilde{T}^2/3\simeq
1.01$, which hardly affects the energy scale.

A zero temperature strongly-interacting gas expands by a
hydrodynamic scale factor $b^H_\perp(t_{heat})$, when released
from a harmonic trap~\cite{OHaraScience,Menotti}. Heating arises
after recapture and subsequent equilibration, but not during
expansion. This follows from the lowest $\tilde{T}=0.04$, obtained
by imaging the gas 1 ms after release from the trap. Hence, the
temperature change during $t_{heat}\leq 460\,\mu$s $< 1$ ms must
be very small.

Thermometry of strongly-interacting Fermi gases is not well
understood. By contrast, thermometry of noninteracting Fermi gases
can be simply accomplished by fitting the spatial distribution of
the  cloud (after release and ballistic expansion)  with a
Thomas-Fermi (T-F) profile, which is a function of two parameters.
We choose them to be the Fermi radius $\sigma_x$ and the reduced
temperature $T/T_F$. However, this method is only precise at
temperatures well below $0.5\, T_F$, where $\sigma_x$ and $T/T_F$
are determined independently. At higher temperatures, where the
Maxwell-Boltzmann limit is approached, such a fit determines only
the product $\sigma_x^2\, T/T_F$. We  circumvent this problem by
determining $\sigma_x$ from a low temperature fit, and then hold
it constant in the fits at all higher temperatures, enabling a
one-parameter determination of the reduced temperature.

Spatial profiles of strongly-interacting Fermi gases closely
resemble T-F distributions, as observed
experimentally~\cite{OHaraScience,Grimmbeta} and as
predicted~\cite{LevinDensity}. The profiles of the trapped and
released gas are related by hydrodynamic scaling to a good
approximation. Over a wide temperature range, this scaling is
consistent with the observed cloud size to $\pm$ 2\%  and is
further supported by measurements of the breathing frequency,
which are within $\pm$ 1\% of the unitary hydrodynamic
value~\cite{Kinast}. Analogous to the noninteracting case, we
define an experimental dimensionless temperature parameter
$\tilde{T}$, which is determined by fitting the cloud profiles
with a T-F distribution~\cite{Jackson}, holding constant the Fermi
radius of the interacting gas, $\sigma_x'$. We find experimentally
that $\tilde{T}$ increases monotonically from the highly
degenerate regime to the Maxwell-Boltzmann limit. This fitting
procedure also leads us to define a natural reduced temperature
scale in terms of the zero temperature parameters $\beta$ and
$T_F$,
\begin{equation}
\tilde{T}_{nat}\equiv\frac{k_BT}{\mu_0}=\frac{T}{T_F\sqrt{1+\beta}}.
\label{eq:temp}
\end{equation}
Eq.~\ref{eq:temp} is consistent with our choice of fixed Fermi
radius $\sigma_x'$, i.e., $m\omega_x^2\sigma_x'^2/2=\mu_0$. At
high temperatures, we must interpret $\tilde{T}=\tilde{T}_{nat}$,
to obtain the correct Maxwell-Boltzmann limit. At low
temperatures, $\tilde{T}\simeq \tilde{T}_{nat}$ yields an estimate
of $T/T_F$ which can be further calibrated to the theoretical
reduced temperature $T/T_F$ by performing the experimental fitting
procedure on the theoretically generated density
profiles~\cite{ChenScience,SupportOnline}.

Preliminary data processing yields normalized, one-dimensional
spatial profiles of the atomic cloud~\cite{SupportOnline}. To
determine $\tilde{T}$ over the full temperature range of interest,
we employ a fixed expansion time of 1 ms. We first measure
$\sigma_x'$ from our lowest temperature data. Then, $\tilde{T}$ is
determined using the one parameter T-F fit method. This yields
$\tilde{T}=0.04-2.15$ for the strongly-interacting gas.

The experimental energy scale Eq.~\ref{eq:E0} and the natural
temperature scale Eq.~\ref{eq:temp} are determined by measuring
the value of $\beta$.  This is accomplished by comparing the
measured radius of the strongly-interacting gas $\sigma_x'$ to the
radius for a noninteracting gas~\cite{SupportOnline}. We find that
$\beta =-0.49(0.04)$ (statistical error only) in reasonable
agreement with the best current predictions, where
$\beta=-0.56$~\cite{Carlson}, and $\beta =-0.545$~\cite{Strinati}.

We now apply our energy input and thermometry methods to measure
the heat capacity of our optically trapped Fermi gas, i.e., for
different values of $t_{heat}$, we measure the temperature
parameter $\tilde{T}$  and calculate the total energy
$E(t_{heat})/E_0$ from Eq.~\ref{eq:energy}. The time $t_{heat}$
determines the energy accurately, as the trap intensity switches
in less than $1\,\mu$s. We believe that shot-to-shot fluctuations
in the energy are negligible, based on the small fractional
fluctuations in $\tilde{T}$ at low temperatures, where the heat
capacity is expected to be very small.  To obtain high resolution
data, 30-40 different heating times $t_{heat}$ are chosen. The
data for each of these heating times are acquired in a random
order to minimize systematic error. Ten complete runs are taken
through the entire random sequence.

We first measure the heat capacity for a noninteracting Fermi
gas\cite{Kinast,SupportOnline}, where the scattering length $a$ is
zero. This occurs near 526 G. Fig. 1
%~\ref{fig:LinTemp}
shows the data (green dots) which represent the calculated
$E(t_{heat})/E_0$ versus the measured value of $\tilde{T}$, for
each $t_{heat}$. For comparison, predictions for a noninteracting,
trapped Fermi gas, $E_{ideal}(\tilde{T})/E_{ideal}(0)$ are shown
as the black curve, where $\tilde{T}=T/T_F$ in this case. Here,
the chemical potential and energy are calculated using a finite
temperature Fermi distribution and the density of states for the
trapped gas. Throughout, we use the density of states for a
realistic Gaussian potential well, $U(r) = U_0[1-e
^{-m\bar{\omega}^2r^2/2U_0}]$ with $U_0=14.6\,k_BT_F$, rather than
the harmonic oscillator approximation. This model is in very good
agreement with the noninteracting gas data  at all temperatures.

For the strongly-interacting gas at 840 G, Fig.~1 (blue diamonds),
the gas is cooled to $\tilde{T}=0.04$ and then heated.
%~\ref{fig:LinTemp}
Note that the temperature parameter $\tilde{T}$ varies by a factor
of 50 and the total energy by a factor of 10. For comparison, we
show the theoretical results for the unitary case as the red
curve. Here the horizontal axis for the theory is obtained using
the approximation $\tilde{T}\simeq \tilde{T}_{nat}$ via
Eq.~\ref{eq:temp}. On a large scale plot, the data for the
strongly-interacting and noninteracting gases appear quite
similar, although there are important differences at low
temperature.
%~\ref{fig:LinTemp}

A striking result is observed by plotting the low temperature data
of Fig.~1 on an expanded scale~\cite{KinastScience,SupportOnline}.
This reveals a transition in the heat capacity which is made
evident by plotting the data for the strongly-interacting gas on a
$log-log$ scale as in Fig. 2. The transition is apparent in  the
raw temperature data~\cite{KinastScience,SupportOnline}, and is
strongly suggestive of the onset of superfluidity. Note that the
observed spatial profiles of the gas vary smoothly and are closely
approximated by T-F shapes in the transition region. Fig. 2 shows
the transition after converting the empirical temperature
$\tilde{T}$ to theoretical $T/T_F$ units.

The empirical temperature is calibrated to enable precise
comparison between the theory and the experimental data. For the
calibration, we subject the theoretically derived density profiles
~\cite{LevinDensity,ChenScience} to the same one-dimensional T-F
fitting procedure as used in the experiments. One dimensional
density distributions are obtained by integrating over two of the
three dimensions of the predicted  spatial profiles, which are
determined for a spherically symmetric trap. Our results for this
temperature calibration are shown in the inset to Fig.~2. This
calibration provides a mapping between the experimental reduced
temperature $\sqrt{1+\beta}\,\tilde{T}$ and the theoretical
temperature $T/T_F$. We find that $\tilde{T}=\tilde{T}_{nat}$ is a
very good approximation above $T_c$.  Such scaling may be a
manifestation of universal
thermodynamics~\cite{HoUniversalThermo}. The difference between
$\tilde{T}$ and $\tilde{T}_{nat}$ is significant only below the
superfluid transition $T_c$ and is therefore negligible in the
large scale plot of Fig.~1 over a broad temperature range.
However, below $T_c$ the fits to the theoretical profiles yield a
value of $\sqrt{1+\beta}\,\tilde{T}$ which is lower than the
theoretical value of $T/T_F$. This is a consequence of condensate
effects~\cite{SupportOnline}.

Fig.~2 shows that above a certain temperature $T_c$, the
strongly-interacting data nearly overlap that of the
noninteracting gas, and exhibit a power law fit
$E/E_0-1=4.98(T/T_F)^{1.43}$. Below $T_c$, the data deviate
significantly from noninteracting Fermi gas behavior, and are well
fit by $E/E_0-1=97.3(T/T_F)^{3.73}$ (dashed curve). From the
intersection point of these power law fits, we estimate
$T_c/T_F=0.27(.02)$ (statistical error only). This is very close
to our theoretical value $T_c/T_F = 0.29$.

The fractional change in the heat capacity $C$ is estimated from
the slope change in the fits to the calibrated data. In that case,
the relative specific heat jump $(C_<-C_>)/C_>\approx 1.51 (0.05)$
(statistical error only), where $>(<)$ denotes above (below)
$T_c$. This is close to the value (1.43) for an $s$-wave BCS
superconductor in a homogeneous case, although one expects
pre-formed pairs, i.e., pseudogap effects, to modify the
discontinuity somewhat~\cite{Chen4}.

In Fig. 2 and Fig. 3, the theory is compared to the calibrated
data after very slightly detuning the magnetic field in the model
away from resonance, so that the predicted unitary gas parameter
$\beta$ has the same value as measured. This small detuning,
$(k_Fa)^{-1}=0.11$, where $k_F=\sqrt{2m\,k_B\,T_F/\hbar^2}$, is
reasonable given the broad Feshbach
resonance~\cite{BartensteinFeshbach} in $^6$Li.

Finally, Fig.~3 presents an expanded view of the low temperature
region. Here, the experimental unitary data is calibrated and
replotted in the more conventional theoretical units, $E_F=k_BT_F$
and $T_F$. The agreement between theory and experiment is very
good. In the presence of a pseudogap, a more elaborate treatment
\cite{Chen4} of the pseudogap self-energy, which takes into
account spectral broadening, will be needed in order to calculate
accurately the specific heat jump.

If one extends the temperature range in Fig.~3
%~\ref{fig:LowT}
to high $T$ we find that both the unitary and noninteracting cases
coincide above a characteristic temperature, $T^*$, although below
$T_c$ they start out with different power laws (as shown in
Fig.~2).
%~\ref{fig:LogTemp}).
In general, we find that agreement between theory and experiment
is very good over the full temperature range for which the data
are taken.  The observation that the interacting and
noninteracting curves do not precisely coincide until temperatures
significantly above $T_c$ is consistent with (although it does not
prove) the existence of a pseudogap and with onset temperature
from the figure $ T^* \approx 2 T_c$.   Related signatures of
pseudogap effects  are also seen in the thermodynamics of high
temperature superconductors~\cite{Levin}.
%Although one might be inclined to
%interpret the data in Fig. 1
%~\ref{fig:LinTemp}
%as reflecting a form of
%universality at low $T$, Figs. 2 and 3
%~\ref{fig:LogTemp} and \ref{fig:LowT}
%(when extended to a wider range of temperatures) shows that the overlap
%indicated in Fig. 1
%~\ref{fig:LinTemp}
%will break down for $ T \gg T^*$.

%\bibliography{ScienceHeatCap,ScienceReview2}

\begin{scilastnote}

\item We thank T.-L. Ho, N. Nygaard, C. Chin, M. Zwierlein, M.
Greiner and D.S. Jin for stimulating correspondence. This research
is supported by the Chemical Sciences, Geosciences and Biosciences
Division of the Office of Basic Energy Sciences, Office of
Science, U. S. Department of Energy, the Physics Divisions of the
Army Research Office and the National Science Foundation, the
Fundamental Physics in Microgravity Research program of the
National Aeronautics and Space Administration, NSF-MRSEC Grant No.
DMR-0213745, and in part by the Institute for Theoretical
Sciences, a joint institute of Notre Dame University and Argonne
National Laboratory and by the U.S. Department of Energy, Office
of Science through contract number W-31-109-ENG-38.

\end{scilastnote}

\noindent \textbf{Supporting Online Material}\\
www.sciencemag.org\\
Materials and Methods\\
Figs. S1, S2\\
Supporting References and Notes

\clearpage
%FIGURE 1
\begin{figure}
\begin{center}\
\includegraphics[]{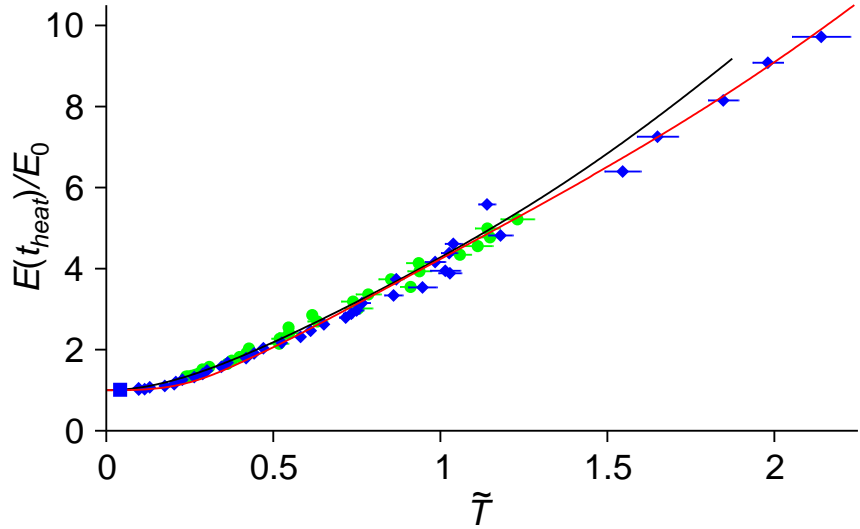}
\end{center}
\caption{Total energy versus temperature. For each heating time
  $t_{heat}$, the temperature parameter $\tilde{T}$ is measured from
  the cloud profile, and the total energy $E(t_{heat})$ is calculated
  from Eq.~(\ref{eq:energy}) in units of the ground state energy
  $E_0$. Green circles: noninteracting Fermi gas data; Blue diamonds:
  strongly-interacting Fermi gas data. Black curve: predicted energy
  versus reduced temperature for a noninteracting, trapped Fermi gas,
  $E_{ideal}(\tilde{T})/E_{ideal}(0)$; Red curve: predicted
  energy versus $\tilde{T}$ for the unitary case. No temperature
  calibration is applied since $\tilde{T}\approx \tilde{T}_{nat}$ over
  the broad temperature range shown.  Note that the lowest temperature
  point (blue square) is constrained to lie on the black curve.}
\label{fig:LinTemp}
\end{figure}

%FIGURE 2
\begin{figure}
\centerline{\includegraphics[width=5.0in,clip]{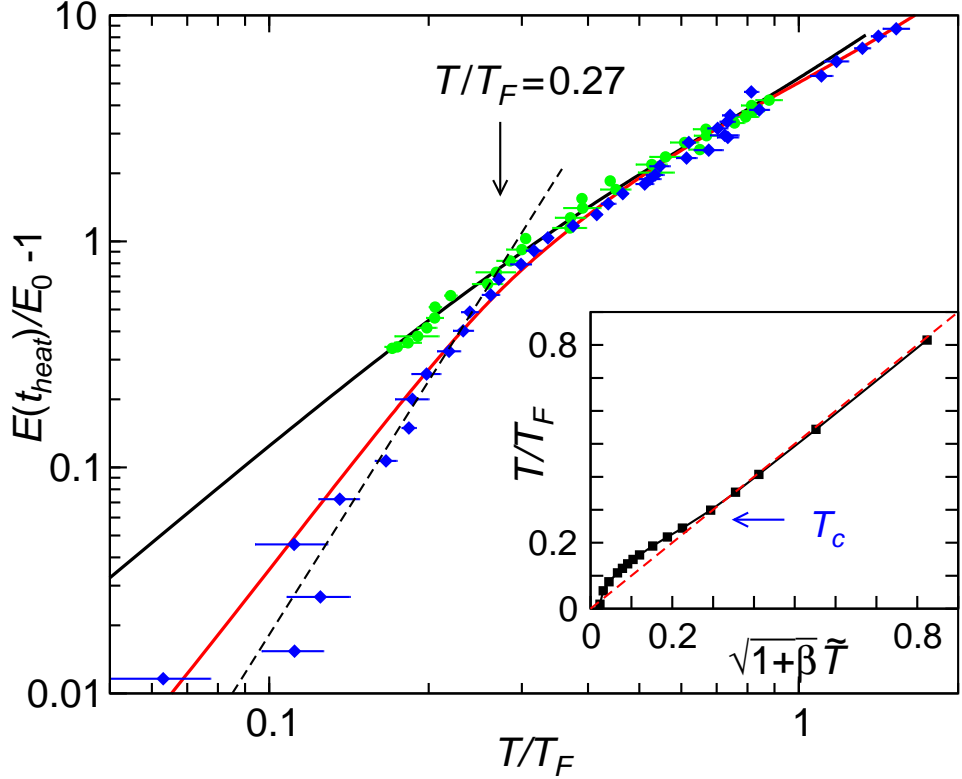}}
\caption{Energy input versus temperature from
  Fig.~1 after temperature calibration on a $log-log$ scale. The
  strongly-interacting Fermi gas shows a transition in behavior near
  $T/T_F=0.27$. Green circles: noninteracting Fermi gas data; Blue
  diamonds: strongly-interacting Fermi gas data; Red (Black) curve:
  prediction for a unitary (noninteracting),  Fermi gas
   in a Gaussian trap as in experiment; Black dashed
  line: best fit power law $97.3\,(T/T_F)^{3.73}$ to the unitary
  data for $T/T_F\leq 0.27$.
% $9.8\,\tilde{T}^{2.53}$.
%$27.7 \tilde{T}_{nat}^{3.73}$
%Note the lowest temperature point
%  (blue square) is not included in the fits, as it is constrained to lie
%  on the red curve.
%Comparison of present theory (lines) and
%  experiments (symbols) of Ref.~\cite{Kinast} in terms of $E/E_0-1$ as a
%  function of $\tilde{T}_{nat}\equiv T/(T_F\sqrt{1+\beta})$ on a log-log
%  scale, for both unitary and noninteracting cases, with Gaussian trap
%  potential having trap depth $V_0/E_F=14.6$ as in experiment.
  The inset shows the calibration curve, which has been applied to the
  unitary data (blue diamonds). The red dashed line in the inset
  represents the diagonal, $T/T_F = \sqrt{1+\beta}\, \tilde{T}$. Here
  $E_0 \equiv E(T=0)$.}
\label{fig:LogTemp}
\end{figure}

%FIGURE 3
\begin{figure}[tb]
\centerline{\includegraphics[width=5.in,clip]{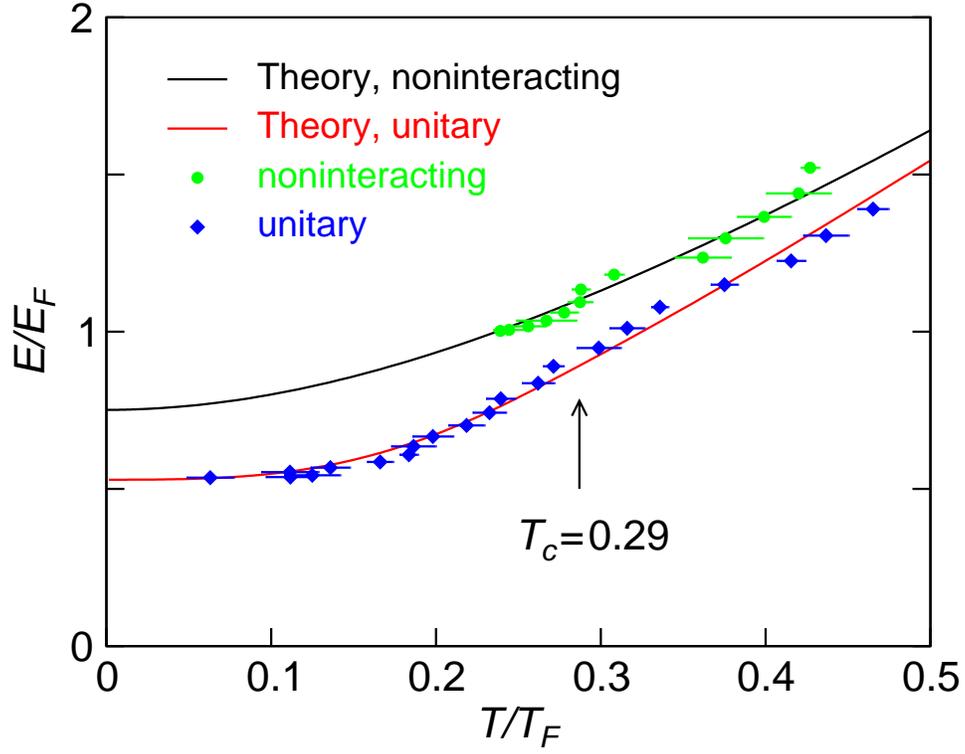}}
\caption{Low temperature comparison of present theory (red, black
curves) and   experiments (symbols) in terms of $E/E_F$
($E_F=k_BT_F$) per atom as a function of   $T/T_F$, for both
unitary and noninteracting gases in a Gaussian trap. The fact that
the two experimental (and the two theoretical) curves do not merge
until higher $T^* > T_c$ is consistent with the presence of a
pseudogap. }
 \label{fig:LowT}
 \end{figure}

% Begin online here

\setcounter{figure}{0} \setcounter{equation}{0}
\renewcommand{\thefigure}{S\arabic{figure}}
\renewcommand{\theequation}{S\arabic{equation}}

\makeatletter \renewcommand\@biblabel[1]{S#1.} \makeatother
\renewcommand\citeform[1]{S#1} % parenthesized numbers [S1-S5]
%

%\topmargin 0.0cm \oddsidemargin 0.2cm \textwidth 16cm \textheight
%21.4cm \footskip 1.0cm

%%%%%%%%%%%%%%%%% END OF PREAMBLE %%%%%%%%%%%%%%%%

\clearpage

\newpage
\renewcommand{\thefigure}{S\arabic{figure}}
\setcounter{figure}{0}

\section*{Supporting Online Material}

%\section*{Supporting Online Material: Materials and Methods}
%\subsection*{Finite Temperature BCS-BEC Crossover Theory}
\subsection*{Computation of Thermodynamical Quantities}

The theoretical community is in the midst of unraveling the nature
of resonantly interacting fermionic superfluids
\cite{SJasonHo,SStrinati4,SCarr,SWilliams,SHeiselberg,STosi,SStringari2,STorma2,SCChin}
with particular emphasis on the strongly interacting Fermi gas
\cite{SOHaraScience}.  In the BCS-BEC crossover picture
\cite{SLevin}, the strongly interacting Fermi gas is intermediate
between the weak coupling BCS and BEC limits.  In addressing the
nature of the excitations from the conventional mean field or
BCS-like ground state~\cite{SLeggett}, our theoretical
calculations help to provide a theoretical calibration of the
experimental thermometry, and elucidate the thermodynamics.

Without doing any calculations one can anticipate a number of
features of thermodynamics in the crossover scenario.  The
excitations are entirely bosonic in the BEC regime, exclusively
fermionic in the BCS regime, and in between both types of
excitation are present.  In the so-called one-channel problem the
``bosons" correspond to noncondensed Cooper pairs, whereas in
two-channel models, these Cooper pairs are strongly hybridized
with the molecular bosons of the closed channel, singlet state.
Below $T_c$ the presence of the condensate leads to a
single-branch bosonic excitation spectrum which, at intermediate
coupling, is predominantly composed of large Cooper pairs. These
latter bosons lead to a pseudogap \cite{SLevin,SJS3} above $T_c$.
Within the conventional mean field ground state, and over the
entire crossover regime \cite{SJS2a} below $T_c$, the bosons with
effective mass $M^*$ have dispersion $\Omega_q = \hbar^2
q^2/2M^*$. This form for the dispersion reflects the absence of
direct boson-boson interactions. In the extreme BEC limit, when
the fermionic degrees of freedom become irrelevant, direct
inter-boson interactions must be accounted for. While our focus in
this paper is on the unitary case, when we refer to ``BEC" we
restrict our attention to the near-unitary BEC regime.

As long as the attractive interactions are stronger than those of
the BCS regime, these noncondensed pairs must show up in
thermodynamics, as must the pseudogap in the fermionic spectrum.
These are two sides of the same coin. Below $T_c$, the fermionic
excitations have dispersion $\Ek = \sqrt{(\ek-\mu)^2 + \Delta^2}$,
where $\ek = \hbar^2 k^2/2m$ and $\mu$ are the atomic kinetic
energy and fermionic chemical potential, respectively.  That this
excitation gap $\Delta$ is non-zero at $T_c$ in the Bogoliubov
quasi-particle spectrum $\Ek$, differentiates the present approach
\cite{SJS2a} from all other schemes which address BCS-BEC
crossover at finite $T$.  The bosons, by contrast, are gapless in
the superfluid phase, due to their vanishing chemical potential.
Within a trap, and in the fermionic regime (for which $\mu > 0$),
the fermionic component will have a strong spatial inhomogeneity
via the spatial variation of the gap.  Thus, in contrast to the
homogeneous case, fermions on the edge of the trap, which have
relatively small or vanishing excitation gaps $\Delta$, will
contribute power law dependences to the thermodynamics.
%Similarly,

Starting at a magnetic field well above a Feshbach resonance, by
decreasing the magnetic field, we tune from the BCS-like regime
towards unitarity at resonance. We first consider low $T$ where
fermions become paired over much of the trap.  The unpaired
fermions at the edge tend to dominate the thermodynamics
associated with the fermionic degrees of freedom, and lead to a
higher (than linear) power law in the $T$ dependence of the
entropy.  The contribution from excited pairs of fermions is
associated with a $T^{3/2}$ dependence of entropy on temperature
which dominates for temperatures $T/T_F \lesssim 0.05$ or $T/T_c
\lesssim 0.2$.  In general, the overall exponent of the low $T$
power law varies with magnetic field, depending on the magnitude
of the gap and temperature, as well as the relative weight of
fermionic and bosonic contributions.  In the superfluid phase, at
all but the lowest temperatures, the fermions and bosons combine
to yield $S \propto T^2$ precisely at resonance ($(k_Fa)^{-1}=0$).
For the near-unitary case investigated in the paper
($(k_Fa)^{-1}=0.11$), we have $ S \propto T^{1.9}$.

Because our calculations \cite{SChenScience} are based on the
standard mean field ground state \cite{SLeggett}, we differ from
other work\cite{SStrinati4,SStrinati5} at finite temperatures.
Elsewhere \cite{SJS2a,SJS3,SLevinDensity} we have characterized in
quantitative detail the characteristic gap $\Delta$ and pseudogap
$\Delta_{pg}$ energy scales.  The pseudogap (which is to be
associated with a hybridized mix of noncondensed fermion pairs and
molecular bosons) and the superfluid condensate (sc) called
$\tilde{\Delta}_{sc}$, add in quadrature to determine the
fermionic excitation spectrum: $\Delta^2(T) =
\tilde{\Delta}_{sc}^2(T) + \Delta_{pg}^2(T)$.  Our past work
\cite{SJS2a,SJS3,SLevinDensity} has primarily focussed below
$T_c$. Here we extend these results, albeit approximately, above
$T_c$. Our formalism has been applied below $T_c$ with some
success in Ref.~\cite{STorma2} to measurements of the pairing gap
in RF spectroscopy.  A more precise, but numerically more complex
method for addressing the normal state was given in
Ref.~\cite{SMaly1}.

After including the trap potential $U(r)$ and internal binding
energy of the bosons, the local energy density can be decomposed
into fermionic ($E_f$) and bosonic ($E_b$) contributions and
directly computed as follows
\begin{eqnarray}
E &=& \mu n(r) + E_f + E_b \,, \nonumber\\
E_f &=& \sum_K (i\omega_n +\ek - \mu(r)) G(K) \nonumber\\
&=& \sumk [2\Ek f(\Ek) -(\Ek-\ek+\mu(r))]
+ \Delta^2 \chi(0) \,, \nonumber\\
E_b &=& \sum_q (\Omegaq-\mu_{boson})\, b(\Omegaq-\mu_{boson}) \,,
\end{eqnarray}
where $\mu(r) = \mu - U(r)$, $n(r)$ is the local density,
$\omega_n = (2n+1)\pi k_B T$ is the fermionic Matsubara frequency,
$G(K)$ is the renormalized fermionic Green's function with
four-momentum $K\equiv (i\omega_n, \mathbf{k})$, $b(x)$ and $f(x)$
are the Bose and Fermi distribution functions, respectively.  The
pair susceptibility $\chi(0)$, at zero frequency and zero
momentum, is given by
\begin{equation}
\chi(0) = \sum_{\bf k}
 \frac{1-2 f(\Ek)}{2 \Ek}
\end{equation}
and the bosonic chemical potential $\mu_{boson}$ is zero in the
superfluid phase.

Unlike the situation in condensed matter systems, for these
ultracold gases, thermometry is less straightforward.
Experimentally, temperature is determined from the spatial
profiles of the cold gas, either in the trap, or following
expansion. For weakly interacting Bose and Fermi gases, where the
theoretical density is well understood, this procedure is
straightforward. However, for a strongly interacting gas, the
spatial profile has not been understood until
recently~\cite{SLevinDensity}. For this reason, the temperature is
often measured on either side far away from the Feshbach
resonance, where the scattering length is small. A strongly
interacting sample in the unitary regime is then prepared by an
adiabatic change of the magnetic field.

More specifically, in the BCS or weak attraction regime,
temperature is determined by fitting the spatial (or momentum
distribution) profiles to those of a non-interacting Fermi gas
\cite{SJin4}.  In the opposite BEC regime, temperature can be
deduced by fitting the Gaussian wings of density profiles or
determining condensate fractions \cite{SGrimm2a,SKetterle3a}.
Thus, it is convenient to describe a given intermediate regime
which is accessed adiabatically, by giving the initial temperature
at either endpoint.  In order to determine this adiabatically
accessed temperature, one needs precise knowledge of the entropy
$S$ as a function of $T$ and magnetic field from BCS to BEC.
The entropy $S$ can be calculated directly \cite{SChenScience} as
a sum of fermionic and bosonic contributions based on the two
types of excitations. Equivalently, one can also calculate the
entropy from the energy, $S=\int_0^T \frac{dT}{T}\frac{dE}{dT} $.

In the strongly interacting regime, one can measure an empirical
temperature $\tilde{T}$ by fitting a T-F density profile directly
to the spatial distribution, as done in this paper. In the
following, we describe a temperature calibration method which
relates the measured empirical temperature $\tilde{T}$ to the
theoretical value of $T/T_F$.

\subsection*{Calibration of Experimental Temperature Scale}

In order to obtain a temperature calibration curve for the
experiments (inset, Fig.~2 main text) we note that our
theoretically generated profiles yield very good agreement with
the Thomas-Fermi functional form \cite{SLevinDensity} for the
normal and superfluid states. However, there are slight systematic
deviations from this form in the superfluid phase. Below $T_c$ the
profiles contain the superfluid condensate as well as
non-condensed pairs along with excited fermions. Although our
profiles are generated for an isotropic trap, it can easily be
shown that trap anisotropy is not relevant for thermodynamic
quantities. Because they involve integrals over the entire trap,
the calculations can be mapped onto an equivalent isotropic
system.

Our theoretical profiles are generated for given reduced
temperatures $T/T_F$. If one applies the experimental procedure to
these theoretical profiles one can deduce the parameter $\sqrt{1+
\beta}\, \tilde T $ for each $T/T_F$.  Theoretically, then, it is
possible to relate these two temperature scales.  This is
summarized by the calibration curve in the inset to Figure 2.

Quite remarkably, it can be seen from this inset that the
experimental T-F fitting procedure yields the precise theoretical
temperature in the normal state. This applies even below the
pseudogap onset temperature $T^*$, since the non-condensed pairs
and the fermions both are thermally distributed.  However, in the
superfluid phase, the parameter $\sqrt{1+ \beta}\, \tilde T $
systematically underestimates the temperature, because of the
presence of a condensate.  One can understand this effect as
arising principally from the fact that the region of the trap
occupied by the condensate is at the center and decreases in
radius as temperature is increased, until it vanishes at $T_c$.
This prevents the profile from expanding with temperature as
rapidly as for the non-interacting fermions of strict T-F theory.
Hence, one infers an apparently lower temperature.  As $T/T_F$
approaches zero, the parameter $\sqrt{1+ \beta}\, \tilde T $ must
approach zero as well.

\subsection*{Experimental Methods and Empirical Thermometry}

Preparation of the strongly interacting Fermi gas is described in
the main text and the details can be found
elsewhere~\cite{SOHaraScience,SThomas2a,SKinastMagDep}.

Preparation of degenerate, noninteracting Fermi gases follows a
similar series of steps.  As described
previously~\cite{SThomas2a}, 23 s of forced evaporation at 300 G
brings the temperature of the gas to $\tilde{T}=0.24$, the lowest
temperature we can achieve in this case.  The gas is then heated
as described in the main text. Finally, the gas is released and
imaged at 526 G to determine the number of atoms and the
temperature. Temperatures $\tilde{T}$ between 0.24 and 1.23 are
obtained for the noninteracting gas.

All heating and release for time of flight measurements are
conducted at 4.6\% of the full trap depth.  At this depth, the
measured trap frequencies, corrected for anharmonicity, are
$\omega_\perp=\sqrt{\omega_x\omega_y} = 2\pi\times 1696(10)$ Hz
and $\omega_z=2\pi\times 72(5)$ Hz, so that
$\bar{\omega}=(\omega_x\omega_y\omega_z)^{1/3}=2\pi\times 592(14)$
Hz is the mean oscillation frequency.

For both the interacting and noninteracting samples, the column
density is obtained by absorption imaging of the expanded cloud
after 1 ms time of flight, using a two-level state-selective
cycling transition~\cite{SOHaraScience,SThomas2a}. In the
measurements, we take optical saturation into account exactly and
arrange to have very small optical pumping out of the two-level
system.  The resulting absorption image of the cloud can then be
analyzed to determine the temperature of the sample.

\subsection*{Anharmonic Corrections to the Energy Input}

Eq.~2 of the main text does not include corrections to the energy
input which arise from anharmonicity in the gaussian beam trapping
potential. In general, after the cloud expands for a time
$t_{heat}$, the energy changes when the trapping potential
$U(\mathbf{x})$ is abruptly restored,
\begin{equation}
\Delta E(t_{heat})=\int
d^3\mathbf{x}[n(\mathbf{x},t_{heat})-n_0(\mathbf{x})]U(\mathbf{x})\,.
\label{eq:Einput}
\end{equation}
Here $n(\mathbf{x},t_{heat})$ ($n_0(\mathbf{x})$) is the density
of the expanded (trapped) cloud, where $n_0(\mathbf{x})$ is a zero
temperature T-F profile, as noted in the main text. A scale
transformation~\cite{SOHaraScience,SMenotti} relates
$n(\mathbf{x},t_{heat})$ to $n_0(\mathbf{x})$. Using this result,
we obtain Eq.~2 of the main text as well as the anharmonic
correction $\Delta E$ arising for a gaussian beam trapping
potential. For a cylindrically symmetric trap, we obtain,
\begin{equation}
\frac{\Delta E}{E_0}
=-\frac{\mu_0}{30\,U_0}\,\left[\,2b_\perp^4(t)+b_\perp^2(t)-3\,\right]+\frac{\mu_0^2}{360\,
U_0^2}\,\left[\,4b_\perp^6(t)+2 b_\perp^4(t)+3
b_\perp^2(t)-9\,\right]. \label{eq:anharmenergy}
\end{equation}
Note that for our experiments, we assume a gaussian beam potential
with three different dimensions.  These corrections are most
significant for the largest values of $t_{heat}$, since the
largest contribution to the energy change arises from atoms at the
edges of the cloud.

\subsection*{Energy Input for Noninteracting Samples}

Although the interacting and noninteracting samples are heated in
the same fashion, there are a few differences in the way the
energy input is calculated.  In the noninteracting case, the
correction factor in Eq. 2 of the main text, $\eta_{nonint}$, is
determined at the lowest temperature $\tilde{T}=0.24$ from the
energy for an ideal Fermi gas. Furthermore, whereas the strongly
interacting gas expands hydrodynamically, expansion of the
noninteracting gas is ballistic so that
$b_\perp(t_{heat})=b^B_\perp(t_{heat})=\sqrt{1+(\omega_\perp
  t_{heat})^2}$.

\subsection*{Determination of $\beta$}

We determine $\beta$ by comparing the measured Fermi radius for
the strongly interacting sample $\sigma_x'$ to the calculated
radius for a noninteracting gas $\sigma_x$ confined in the same
potential. The relation is given by
$\sigma'_x=\sigma_x(1+\beta)^{1/4}$~\cite{SMechStab}, where
$\sigma_x=\sqrt{2k_BT_F/(M\omega_x^2)}$ is the radius for a
noninteracting gas. We obtain $\sigma_x=1.065\, (N/2)^{1/6}\,\mu$m
for our trap parameters. This calculated radius is consistent with
the value measured for noninteracting samples at 526 G in our
trap.  To determine $\sigma_x'$, we measure the size of the cloud
after 1 ms of expansion, and scale it down by the known
hydrodynamic expansion factor of
$b^H_\perp(1\,\mbox{ms})=13.3$~\cite{SOHaraScience,SMenotti}. We
then determine the Fermi radius
$\sigma_x'=11.98\,(N/2)^{1/6}\,\mu\mbox{m}/13.3=0.901(0.021)\,(N/2)^{1/6}\mu$m.
With these results, we obtain $\beta =-0.49(0.04)$ (statistical
error only).

\subsection*{Observed Transition in Energy versus Empirical
  Temperature $\tilde{T}$}

\begin{figure}
\begin{center}\
\includegraphics[bb= 65 241 506 549]{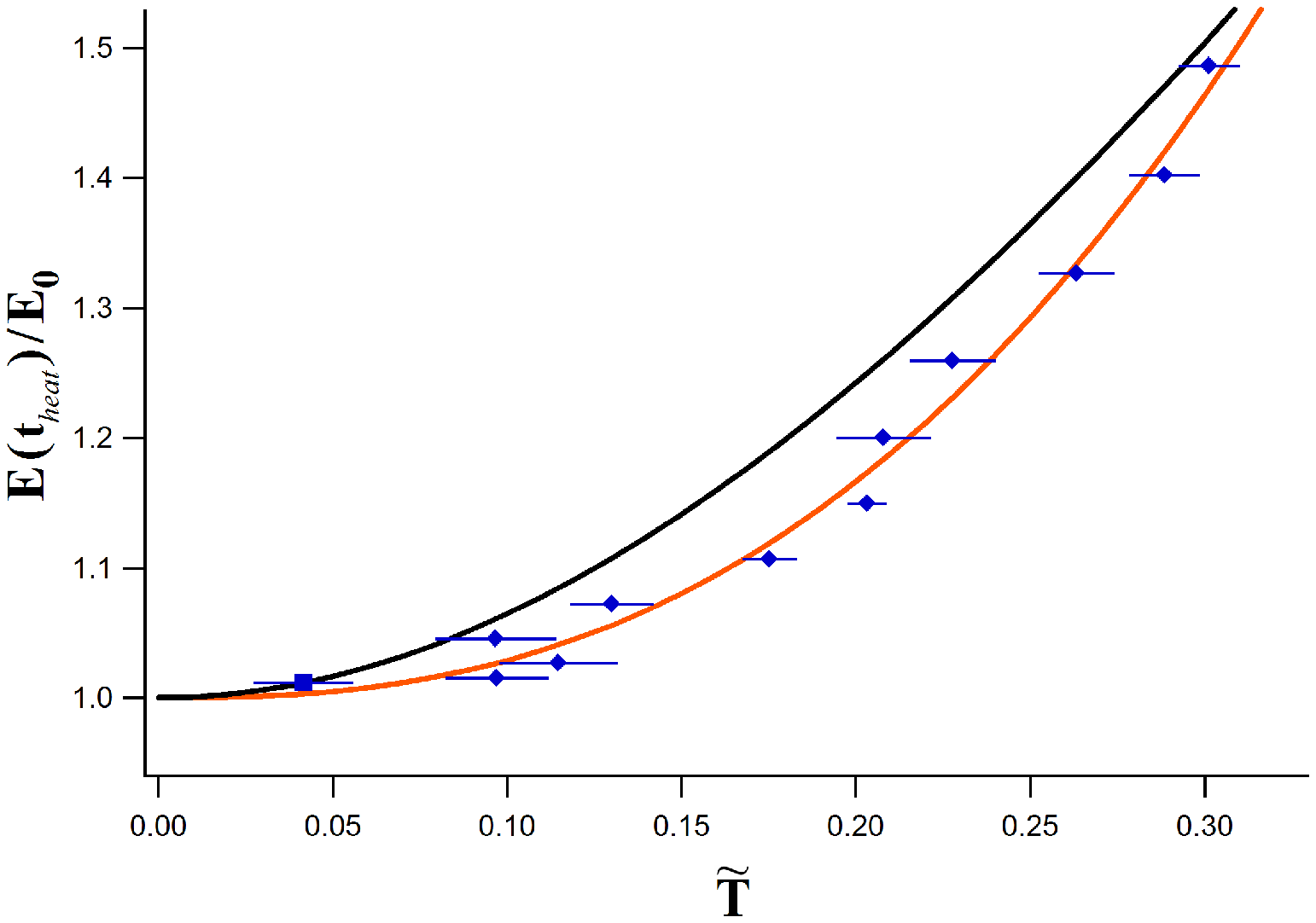}
\end{center}
\caption{Strongly-interacting Fermi gas below the transition
temperature.  $E/E_0$ versus uncalibrated empirical temperature
$\tilde{T}$ on a linear scale. Orange line, best fit power law
$9.8\, \tilde{T}^{2.53}$. Black curve: Predicted $E/E_0$ for an
ideal Fermi gas as a function of $\tilde{T}=T/T_F$. Note the
lowest temperature point (blue square) is not included in the
fits: It is constrained to lie on the black curve by our choice of
$\eta_{int} = 1.01$ in Eq. 2 of the main text.
\label{fig:LowTemp}}
\end{figure}

\begin{figure}
\scalebox{1}{\includegraphics[bb= 85 222 481 560]{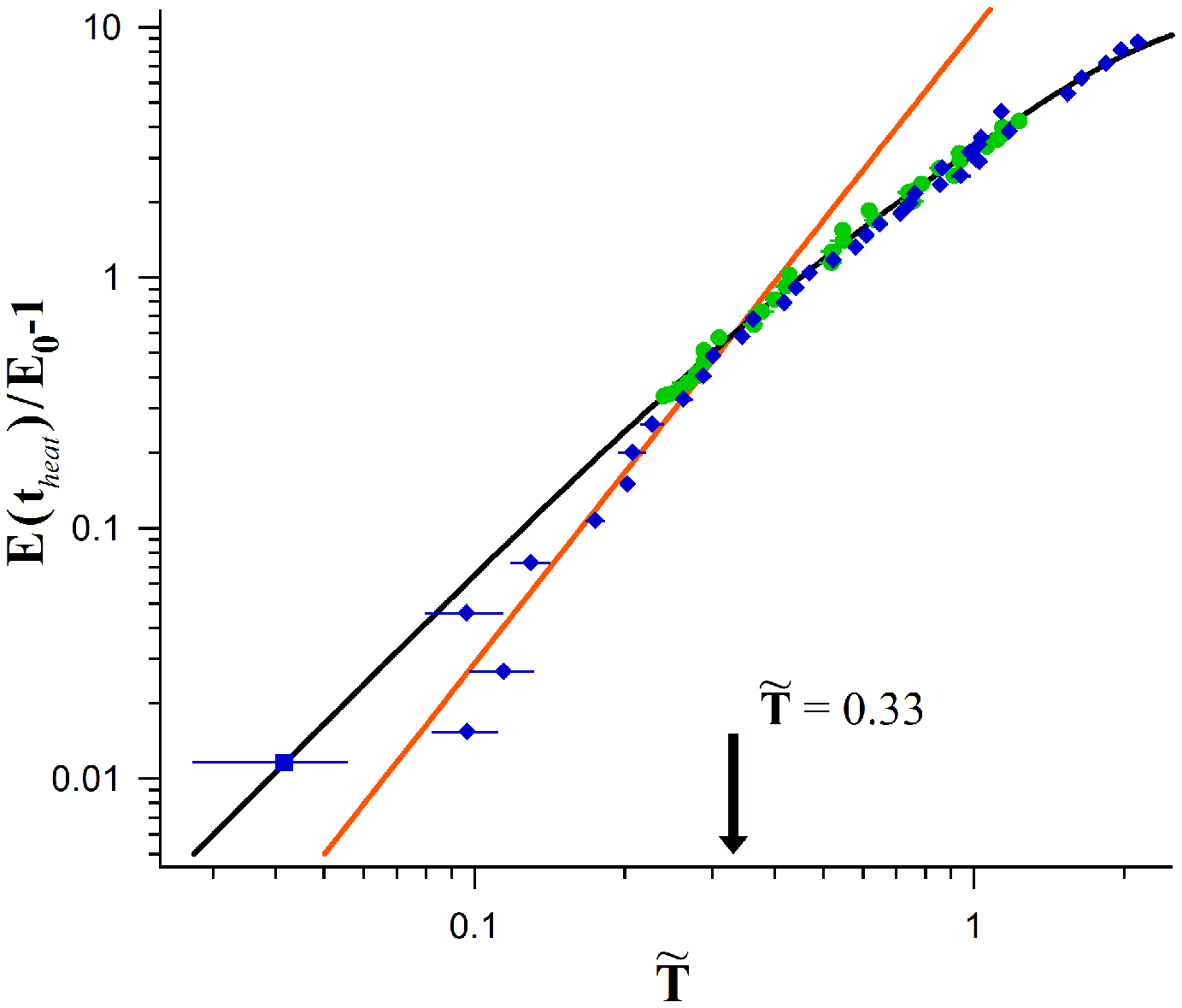}}
%\begin{figure}[tbh]
%\centerline{\includegraphics[width=5in,clip]{LogTemp.eps}}
\caption{Energy input versus uncalibrated temperature $\tilde{T}$
 on a $log-log$ scale. The strongly interacting Fermi gas
shows a transition in behavior near $\tilde{T}=0.33$. Green
circles: noninteracting Fermi gas data; Blue diamonds: strongly
interacting Fermi gas data. Black curve, prediction for a
noninteracting, trapped Fermi gas. Orange line, best fit power law
$9.8\,\tilde{T}^{2.53}$. Note the lowest temperature point (blue
square) is not included in the fits, as it is constrained to lie
on the black curve. } \label{fig:LogUncal}
\end{figure}

For the strongly interacting Fermi gas, without calibrating the
empirical temperature scale, we observe a transition between two
patterns of behavior at $\tilde{T}=0.33$~\cite{SKinastScience}:
For $\tilde{T}=0.33-2.15$, we find that the energy closely
corresponds to that of a trapped Fermi gas of noninteracting atoms
with the mass scaled by $1/(1+\beta)$.  At temperatures between
$\tilde{T}=0.04-0.33$, the energy scales as $\tilde{T}^{2.53}$,
significantly deviating from ideal gas behavior as can be seen in
Fig.~\ref{fig:LowTemp}. The transition between two power laws is
evident in the slope change of the  $log-log$ plot of
Fig.~\ref{fig:LogUncal}.

\clearpage

\clearpage
%\bibliography{ScienceReview2,ScienceHeatCap}

%Figure S1
%

%Figure S1

\end{document}